 \documentclass[prd,preprintnumbers,amsmath,amssymb,floatfix]{revtex4}

\usepackage{graphicx}
\usepackage{dcolumn}
\usepackage{bm}
\usepackage{amssymb}
\usepackage{epsfig}
\usepackage{color}

\newcommand{\ba}{\begin{eqnarray}}
\newcommand{\ea}{\end{eqnarray}}
\newcommand{\be}{\begin{equation}}
\newcommand{\ee}{\end{equation}}
\newcommand{\bdisplay}{\begin{displaymath}}
\newcommand{\edisplay}{\end{displaymath}}

\newcommand{\eq}[1]{Eq.\,(\ref{#1})}

\begin{document}

\title{ Addendum to: ``A new numerical method for obtaining gluon distribution functions $G(x,Q^2)=xg(x,Q^2)$, from the proton structure function $F_2^{\gamma p}(x,Q^2)$.''}  

\author{Martin~M.~Block}
\affiliation{Department of Physics and Astronomy, Northwestern University, 
Evanston, IL 60208}

\date{\today}

\begin{abstract}
In a recent Letter entitled ``A new numerical method for obtaining gluon distribution functions $G(x,Q^2)=xg(x,Q^2)$, from the proton structure function $F_2^{\gamma p}(x,Q^2)$'' \cite{inverseLaplace1}, we derived an accurate and fast algorithm for numerically inverting Laplace transforms, which we used in obtaining gluon distributions from the proton structure function $F_2^{\gamma p}(x,Q^2)$. We inverted the function $g(s)$, where $s$ is the variable in Laplace space, to $G(v)$, where $v$ is the variable in ordinary space. Since publication, we have discovered that the algorithm does not work if $g(s)\rightarrow 0$ less rapidly than $1/s$, as $s\rightarrow\infty$. Although we require that $g(s)\rightarrow 0$ as $s\rightarrow\infty$, it can approach 0 as ${1\over s^\beta}$, with $0<\beta<1$, and still be a proper Laplace transform.  In this note, we derive a new numerical algorithm for just such cases, and test it for $g(s)={\sqrt \pi\over \sqrt s} $, the  Laplace transform of ${1\over \sqrt v}$.
\end{abstract}

\maketitle


\section{Introduction}
In an earlier note \cite{inverseLaplace1} we developed an algorithm to numerically invert Laplace transforms, in order to find an analytic solution for gluon distributions.
We used a global parameterization of the proton structure function, $F_s^{\gamma p}(x,Q^2)$  and  a LO (leading-order) evolution equation for $F_2^{\gamma p}$. 
Subsequently, going to NLO (next-to-leading order) in the strong coupling constant $\alpha_s(Q^2)$,  we have discovered the  algorithm failed badly.  Detailed investigation showed  that the cause of the problem was that for this $g(s)$--whose Laplace transform was our desired NLO  gluon distribution $G(v)$---it went to 0 {\em slower}  than $1/s$ as $s\rightarrow \infty$, where $s$ is the Laplace space variable.  The purpose of this note is to derive a new algorithm for such cases, not covered in \cite{inverseLaplace1}, which can be modeled by Laplace transforms of the type
\ba
g(s)={1\over s^\beta},\qquad 0<\beta<1.\label{stothebeta} 
\ea
To illustrate the new method, we will numerically invert  $g(s)={\sqrt \pi\over \sqrt s} $, the  Laplace transform of ${1\over \sqrt v}$, and test its accuracy, as well as show the inadequacy of the original algorithm for such a case.

\section{Numerical inversion of Laplace transforms}\label{sec:inverseLaplace}

Let $g(s)$ be the Laplace transform of $G(v)$. The Bromwich inversion formula for $G(v)$ is given by
\ba
G(v)\equiv {\cal L}^{-1}[g(s);v]={1\over 2\pi i} \int^{\,c+i\,\infty}_{\,c-i\,\infty}g(s)e^{vs}\,ds,\label{Bromwich}
\ea
 where the real constant $c$ is to the right of all singularities of $g(s)$. 
Making an appropriate coordinate  translation in $s$ so that $c=0$, 
we can write
\ba
G(v)\equiv {\cal L}^{-1}[g(s);v]={1\over 2\pi i} \int^{\,+i\,\infty}_{\,-i\,\infty}g(s)e^{vs}\,ds.\label{Bromwich2}
\ea
 Our goal is to {\em numerically} solve \eq{Bromwich2}. The inverse Laplace transform is essentially determined by the behavior of $g(s)$  near its  singularities, and thus is an ill-conditioned or ill-posed numerical problem. We suggest in this note a new algorithm that takes advantage of very fast, arbitrarily  high precision complex number arithmetic that is possible today in programs like {\em Mathematica} \cite{Mathematica}, making the inversion problem numerically tractable. 

First, we introduce a new complex variable $z\equiv vs$ and rewrite \eq{Bromwich2} as
\ba
 G(v)&=&{1\over 2\pi i v} \int^{\,+i\,\infty}_{\,-i\,\infty}g\left({z\over v}\right)e^{z}\,dz.\label{Bromwichtransformed}
\ea

In our earlier paper \cite{inverseLaplace1}, we proceeded to make a rational approximation to the exponential $e^z$, under the tacit assumption that $g(s)$ went to 0 sufficiently rapidly as $s$ went to $\infty$.  This is {\em not} a valid assumption if $g(s)$ goes to 0 as $1/x^\beta$, $0<\beta<1$. In this case, we will rewrite \eq{Bromwichtransformed} as
\ba
 G(v)&=&{1\over 2\pi i v} \int^{\,+i\,\infty}_{\,-i\,\infty}\left(\frac{g\left({z\over v}\right)}{z^2}\right)\,z^2e^{z}\,dz\label{Bromwichtransformed2}
\ea
and make a rational approximation to  $z^2 e^{z}$,  using  the Pad\'e approximant 
\ba
z^2e^z\approx {\rm Pade}[ze^z,0,(2N-3,2N)]\label{exp_expansion},
\ea
where the numerator of the function Pade in \eq{exp_expansion} is a polynomial in $z$ of order $2N-3$ and whose denominator is a polynomial in $z$  of order  $2N$, and the expansion is around $z=0$.   
Let $\alpha_i,\ i=1,2,\ldots,2N$ be the  $2N$ complex roots of the denominator, i.e., its $2N$ complex poles, and let $\omega_i$ be the complex residues  of the Pad\'e. It can be shown that they have the following properties:
\begin{enumerate}
  \item $\rm{Re}\ \alpha_1>0$, so that its poles are all in the right-hand half of the complex plane. 
  \item $N$ distinct complex conjugate pairs of the complex numbers $(\omega_i,\alpha_i)$, such that the sum of the $k^{\rm th}$ pair,
 \be{\omega_k\over z- \alpha_k}+{\bar \omega_k\over z- \bar\alpha_k},
\ee
 is real for all real $z$. 
 \item The integrand vanishes {\em faster} than $1/R$ as $R\rightarrow\infty$ on the   semi-circle of radius $R$ that encloses the right hand half of the complex plane, since the approximation vanishes as $1/R$ and $g(s)$ that corresponds to a non-singular $ G(v)$ also vanishes for $R\rightarrow\infty$.  
\end{enumerate}
Since g(z/v)  must vanish for $z\rightarrow\infty$  and our approximation      for $z^2e^z$ in \eq{exp_expansion} {\em vanishes} for $z\rightarrow\infty$, we can form a  closed contour $C$ by completing  our integration path of the modified  Bromwich integral in \eq{Bromwichtransformed} with an infinite half circle to the {\em right} half of the complex plane. As mentioned earlier,  $g(z/v)$ has no singularities in this half of the complex-Z plane. It is important to note that this contour is  a {\em clockwise} path  around the poles of \eq{exp_expansion}, which come  from  our approximation to $ze^z$. What we need is the negative of it, i.e., the contour $-C$ which is counterclockwise, so that the poles are to our left as we traverse the contour $-C$. 
Therefore, we rewrite \eq{Bromwichtransformed} as 
\ba
 G(v)&\approx& {1\over 2\pi i v} \oint_C \frac{g\left({z\over v}\right)}{z^2} \rm Pade[z^2e^z,0,(2M-3,2M)]\nonumber\\
&=&-{1\over 2\pi i v} \oint_{-C} \frac{g\left({z\over v}\right)}{z^2} \rm Pade[z^2e^z,0,(2M-3,2M)]\nonumber\\
&= &-{2\over  v} \sum^{N}_{i=1}{\rm Re}\left[\omega_i\frac{g\left({\alpha_i/ v}\right)}{\alpha_i^2}\right].\label{Bromwich3}
\ea
To obtain \eq{Bromwich3}, the final approximation to $ G(v)$,   we used Cauchy's theorem  to equate the closed contour integral around the path $-C$ to $2\pi i$ times the sum of the (complex) residues of the poles.  Since the contour $-C$ restricts us to the right-hand half of the complex plane, no poles of $g(z/v)$ were enclosed, but only the $2N$ poles $\alpha_i$ of the Pad\'e approximant of $z^2e^z$. Using the  properties cited above of the  complex conjugate pairs---$(\omega_i,\alpha_i)$ and  $(\bar \omega_i ,\bar \alpha_i)$---after taking only their real part and multiplying by 2--we have simultaneously insured that $ G(v)$ is real , yet only have had to sum over half of the residues. 
The above scheme now works for the slowly converging case of $g(s)=1/s^\beta$, where $0<\beta<1$, because we are {\em dividing} it by $z^2$ to make the quantity that we are inverting , i.e., $g(z/v)/z$, go to 0 {\em more rapidly} than $1/s$ as $s\rightarrow0$.

Two properties of \eq{Bromwich3} are worth emphasizing here:

\begin{enumerate} 
  \item The $4N$ coefficients ($\alpha_i,\omega_i)$ are  complex constants that are  independent of $v$, only  depending on the value of $2N$ used for the approximation, so that for a given $2N$, they need  to be evaluated only once---in essence, they can be tabulated and stored for later use.  
  
\item 
The real parts of the residues  $\omega_i$  alternate in sign and are {\em exceedingly} large---even for relatively modest $2N$, making round-off a potentially serious problem.  Thus, exceedingly high precision complex arithmetic is called for, often requiring  60 or more digits.  However, this is not a serious problem---either in speed or complexity of execution---for an algorithm written in a program such as {\em Mathematica} \cite{Mathematica}.
\end{enumerate}

A concise inversion algorithm in {\em  Mathematica} that implements \eq{Bromwich3} is  given in Appendix \ref{subsection:inversionalgorithm}.

We will now test  the accuracy of our numerical Laplace inversion algorithm by using $g(s)={\sqrt \pi\over \sqrt s}$, the Laplace transform of $G(v)=\sqrt{v}$.


\section{Comparison of exact solution and numerical Laplace inversion results}
\begin{figure}[h,t,b] 
\begin{center}
\mbox{\epsfig{file=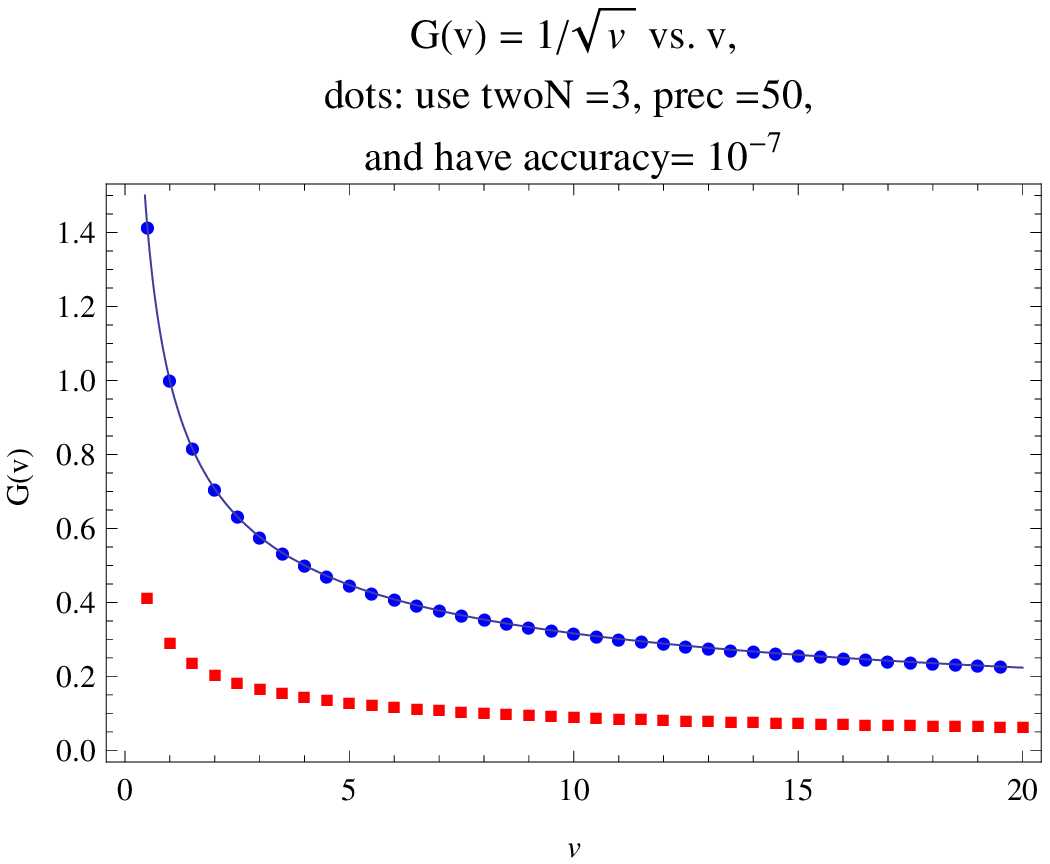
,width=6in%
,bbllx=0pt,bblly=0pt,bburx=420pt,bbury=198pt,clip=%
}}
\end{center}
\caption[]{A plot of $G(v)={1\over \sqrt v}$ vs. v. The solid curve is $G(v)$. The (blue) dots are the approximate values calculated from \eq{Bromwich3},
using  $2N$ =20 and $g(s)={\sqrt \pi\over \sqrt s}$; the (red) squares are the result of using $2N=20$ in the algorithm described in Ref. \cite{inverseLaplace1}, which should {\em only} be used for Laplace transforms that  go to 0 at $\infty$ as fast, or faster than $1/s$.  
\label{fig:Gofv}}
\end{figure}

To illustrate our inversion routine for slowly converging Laplace transforms, we plot in Fig. \ref{fig:Gofv} the numerical inversion of $g(s)={\sqrt \pi\over \sqrt s}$, the Laplace transform of $G(v)={1\over \sqrt v}$.   The solid curve is the function $G(v)={1\over \sqrt v}$, and the (blue) dots are a result of using the algorithm in Appendix \ref{subsection:inversionalgorithm}, using Mathematica \cite{Mathematica}, with $2N=20$.  As seen, the agreement is excellent. Also shown in Fig. \ref{fig:Gofv} are the (red) squares, which are the result of our using our earlier algorithm \cite{inverseLaplace1} for the calculation of the inverse Laplace transform. Clearly, this is a disaster. As detailed in our earlier work \cite{inverseLaplace1}, the original algorithm was {\em exact} if the function in $v$ space was a polynomial of degree $\le 4N-1$, so that it is a very powerful tool when used properly, i.e., with a $g(s)$ that goes to 0 as fast or faster than $1/s$ as $s\rightarrow\infty$. However, for $g(s)$'s that don't vanish this rapidly, that algorithm is  deficient and must be replaced by the new algorithm given in this note.  

As a practical matter, if we have a $g(s)$ to invert that is purely numerical and whose behavior at $\infty$ is completely unknown (except for the requirement that it {\em must} vanish at $\infty$), we must test its convergence properties by  trying {\em both} algorithms.  If they agree reasonably numerically, we should continue using the more powerful algorithm of Ref. \cite{inverseLaplace1}. It they disagree completely, as is the case for the squares  in Fig. \ref{fig:Gofv}, then one should use the new algorithm developed in this note. 

We will discuss in Appendix \ref{subsection:inversionalgorithm} the inherent accuracy of the  algorithm used to get the dots in Fig. \ref{fig:Gofv}.
As noted in  \cite{inverseLaplace1}, our inversion routine 
clearly has a wide variety of additional applications in solving both integral and differential equations. 

\section{Conclusions}
We have achieved arbitrarily high accuracy in numerically  inverting Laplace transforms of the type $g(s)={1\over s^\beta},\ 0<\beta<1$.  As an example, we show in Fig. \ref{fig:Gofv} a comparison of the solution $G(v)={1\over \sqrt v}$, i.e., the inverse Laplace transform of $g(s)={\sqrt \pi\over \sqrt s}$, with the exact answer, and demonstrate the algorithm's inherent  accuracy in Fig. \ref{fig:accuracy}, showing that it is a power law in ${1\over 2N}$, the reciprocal of the number of terms used in the expansion, and thus, of arbitrary accuracy.  Because the assumed $g(s)$ go to 0 very slowly as $s\rightarrow\infty$, the algorithm that we previously developed in Ref. \cite{inverseLaplace1} does {\em not} work here, and we discuss when and when not to use it. We are currently using 
 our new algorithm  (this communication) to analytic decoupled solutions for LO singlet structure functions and gluons \cite{bdhm0} , as well as for obtaining  NLO gluon distributions.
\appendix 
\section{{\em Mathematica} Laplace  Inversion Algorithm}\label{subsection:inversionalgorithm}
\texttt{
NInverseLaplaceTransformBlock2[g\_,s\_,v\_,twoN\_,prec\_]:=Module[\\
\hphantom{xxxxx}\{Omega,Alpha,M,p,den,r,num\},\\
\hphantom{xxxxx}(M=2*Ceiling[twoN/2];If[M<6,M=6];p=PadeApproximant[ z{\^\ }\!\!\!2*Exp[z],\{z,0,\{M-3
,M\}\}];\\
\hphantom{xxxxx}den=Denominator[p];r=Roots[den==0,z];
Alpha=Table[r[[i,2]],\{i,1,M\}];\\
\hphantom{xxxxx}num=Numerator[p];hospital=num/D[den,z];\\
\hphantom{xxxxx}Omega=SetPrecision[Table[hospital/.z->Alpha[[i]],\{i,1,M,2\}],prec+50];\\
 \hphantom{xxxxx}Alpha=SetPrecision[Table[Alpha[[i]],\{i,1,M,2\}],prec]);\\
\hphantom{xxxxx}SetPrecision[-(2/v)Sum[Re[Omega[[i]] (g/(v*s)\^ \ \!\!\!2)/.s->Alpha[[i]]/v],\{i,1,M/2\}],prec]
\\
\hphantom{xxxxxxxxxxxxxxxxxxxxxxxxxxxxxxxxxxxxxxXXXXxxxxxxxxxxxxxxxxxxx}]}

In the above algorithm, \texttt{g} = $g(s)$, \texttt{s} = $s$, \texttt{v} = $v$, \texttt{twoN}=$2N$ in \eq{Bromwich3}, and \texttt{prec} = desired precision of calculation. Typical values are \texttt{twoN}  = 10 and 
\texttt{prec} = 70.  The algorithm, which is quite fast, returns the numerical value of $G(v)$.

The algorithm first insures that  \texttt{M=twoN} is {\em even} and $\ge 6$.  It then constructs \texttt{p}, the Pad\'e approximant whose numerator is a polynomial in $z$  of order $M-3$ and denominator a polynomial in $z$  of order $M$, whose expansion is around $z=0$.  It finds \texttt{r}, the complex roots of the denominator, which are $\alpha_i$, the poles  of \eq{Bromwich3}.  Using L'Hospital's rule, it finds the residue $\omega_i$ corresponding to the pole $\alpha_i$. At this point, all of the mathematics  is symbolic. It next finds every {\em other} pair of $(\alpha_i,\omega_i)$ to the desired numerical accuracy; they come consecutively, i.e., $\alpha_1=\bar \alpha_2,\ \omega_1 = \bar \omega_2,\ \alpha_3=\bar \alpha_4,\ \omega_3 = \bar \omega_4$, etc.  Finally, it  takes the necessary sums, again to the desired numerical accuracy, but only  over  half of the interval $i=1,3,\ldots,N$, by taking  only the real part and multiplying by 2.

\begin{figure}[h,t,b] 
\begin{center}
\mbox{\epsfig{file=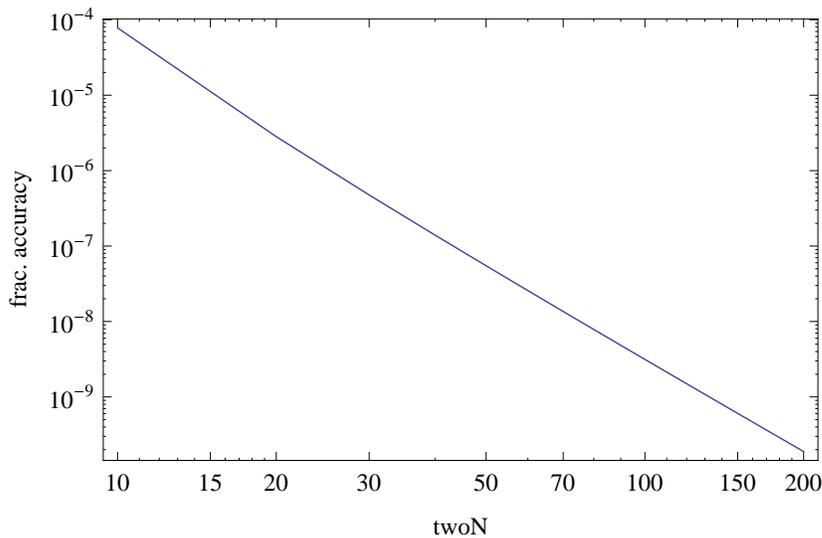
,width=6in%
,bbllx=0pt,bblly=0pt,bburx=420pt,bbury=198pt,clip=%
}}
\end{center}
\caption[]{A log-log plot of the fractional accuracy of the numerical Laplace inversion of the original function $G(v)={1\over \sqrt v}$ vs. twoN, with twoN=$2N$,  the order of the rational approximation in \eq{Bromwich3}. 
\label{fig:accuracy}}
\end{figure}

If $g(s)$, the input to the algorithm, is an {\em analytic} relation {\em and} $v$ is a pure number (from the point of view of {\em Mathematica}, 31/10 is a pure number, but 3.1 is {\em not}), then, for sufficiently high values of {\tt prec}, you can achieve arbitrarily high accuracy. If we define the fractional accuracy as $1- G(v)_{\rm numerical}/ G(v)_{\rm true}$, numerical tests on many different functions shows that it goes to 0 for large {\tt 2N} as a power law in 1/{\tt 2N}.  We illustrate this in Fig. \ref{fig:accuracy}, where we show a log-log plot of the fractional accuracy vs. $2N$.  The straight line shows that the fractional accuracy is a power law in $2N$, allowing one to achieve arbitrary accuracy in the inversion of $g(s)={\sqrt \pi \over \sqrt s}$.  Further discussion of applications and numerical properties can be found in Ref. \cite{inverseLaplace1}.

\begin{acknowledgments}

{\em Acknowledgments:} The author would like to thank the Aspen Center for Physics for its hospitality during the time parts of this work were done.  

\end{acknowledgments}

\bibliography{gluonsPRD.bib}

\end{document}